# Two-dimensional ground-state mapping of a Mott–Hubbard system in a flexible field-effect device

Yoshitaka Kawasugi[1]*†, Kazuhiro Seki[1,2,3]†, Satoshi Tajima[4], Jiang Pu[5], Taishi Takenobu[5], Seiji Yunoki[1,3,6], Hiroshi M. Yamamoto[1,7]*, and Reizo Kato[1]

[1]RIKEN, Wako, Saitama 351-0198, Japan

[2]SISSA - International School for Advanced Studies, Via Bonomea 265, 34136 Trieste, Italy

[3]RIKEN Center for Computational Science (R-CCS), Kobe, Hyogo 650-0047, Japan

[4]Department of Physics, Toho University, Funabashi, Chiba 274-8510, Japan

[5]Department of Applied Physics, Nagoya University, Furo-cho, Chikusa-ku, Nagoya 464-8603, Japan

[6]RIKEN Center for Emergent Matter Science (CEMS), Wako, Saitama 351-0198, Japan

[7]Research Center of Integrative Molecular Systems (CIMoS), Institute for Molecular Science, National Institutes of Natural Sciences, Okazaki, Aichi 444-8585, Japan

*Corresponding author. Email: kawasugi@riken.jp (Y.K.); yhiroshi@ims.ac.jp (H.M.Y.)

†These authors contributed equally to this work.

**Abstract**

A Mott insulator sometimes induces unconventional superconductivity in its neighbors when doped and/or pressurized. Because the phase diagram should be strongly related to the microscopic mechanism of the superconductivity, it is important to obtain the global phase diagram surrounding the Mott insulating state. However, the parameter available for controlling the ground state of most Mott insulating materials is one-dimensional owing to technical limitations. Here we present a two-dimensional ground-state mapping for a Mott insulator using an organic field-effect device by simultaneously tuning the bandwidth and bandfilling. The observed phase diagram showed many unexpected features such as an abrupt first-order superconducting transition under electron doping, a recurrent insulating phase in the heavily electron-doped region, and a nearly constant superconducting transition temperature in a wide parameter range. These results are expected to contribute toward elucidating one of the standard solutions for the Mott–Hubbard model. Theoretical calculations reproducing the main features of our experimental results have also been conducted.

**Introduction**

The electron correlation in solids, or the Mott physics, offers intriguing phenomena in materials such as unconventional superconductivity. The key parameters for controlling the Mott physics are the electronic bandfilling and bandwidth. Variation of the former triggers the doping-induced high-transition-temperature (high-$T_C$) superconductivity in cuprates *(1)* and the field-induced superconductivity in a twisted graphene bilayer *(2)*, while the variation of the latter triggers the pressure-induced superconductivity in organics *(3)* and the chemical-pressure-induced superconductivity in fullerenes *(4)*. However, the simultaneous control of these two parameters has been lacking so far, leaving a comprehensive phase diagram of correlated materials inaccessible. For example, high-$T_C$ cuprates can exhibit superconductivity only in the bandfilling-controlled regime, and the bandwidth cannot be sufficiently controlled to induce superconductivity from a nondoped Mott insulating state because of the hard crystal lattice. Recently, however, we have achieved both electric-field-induced superconductivity using a solid-gate field-effect transistor and strain-induced superconductivity by substrate bending in similar organic Mott insulator devices *(5,6)*. This situation has motivated us to elucidate the two-dimensional ground-state map of the Mott–Hubbard model by combining these two technologies in a single organic device. Here, we report simultaneous control of the bandfilling and bandwidth at an organic Mott insulator interface, where the details of the superconducting transitions in the proximity of the Mott insulator in the two-dimensional ground-state map are revealed for the first time.

The two-dimensional organic Mott insulator κ-(BEDT-TTF)$_2$Cu[N(CN)$_2$]Cl (hereafter κ-Cl) comprises alternating layers of conducting BEDT-TTF$^{+0.5}$ radical cations and insulating Cu[N(CN)$_2$]Cl$^-$ counteranions *(7)*. The conducting BEDT-TTF molecules are strongly dimerized and can be modeled as a single-band Hubbard model on an anisotropic triangular lattice with $t'/t = -0.44$, where $t$ is the nearest-neighbor (interdimer) hopping and $t'$ is the next-nearest-neighbor hopping (Fig. 1A) *(8)*. Similarities between the κ-type BEDT-TTF salts and high-$T_C$ cuprates, such as the proximity between antiferromagnetism and superconductivity, and the unconventional metallic phase in the normal state, have been pointed out *(9)*. In contrast to the high-$T_C$ cuprates, where the superconductivity is induced with carrier doping by chemical substitution, κ-Cl has been studied in terms of the pressure-induced superconducting transition because of its sensitivity to pressure and the difficulty of chemical doping. However, recently developed techniques for field-effect doping have enabled the reversible and finely tuned doping to κ-Cl *(10,11)*. In previous work *(11)*, measurements of transport properties and calculations based on cluster-perturbation theory indicate strong doping asymmetry, where major pseudogaps open near the van Hove critical points under substantial hole doping, while a more non-interacting-like Fermi surface appears under substantial electron doping. The doping asymmetry for the superconductivity, which should be observed under pressure at lower temperatures, is currently of great interest since it will provide further insight into unconventional superconductivities in strongly correlated systems. Here, we report a two-dimensional mapping of the ground state for a Hubbard system based on an organic field-effect device. This is, to our knowledge, the first direct derivation of a ground-state phase diagram with fine control of the bandfilling and bandwidth in a Hubbard system in the solid state (Fig. 1B).

To this end, we fabricated electric-double-layer transistors (EDLTs) using thin single crystals of κ-Cl on flexible polyethylene terephthalate (PET) substrates (Fig. 1C, Fig. S1). EDLT doping is particularly effective for materials with low carrier density because a small gate voltage can markedly vary the bandfilling in such materials. The carrier density of κ-Cl is much lower than that of high-$T_C$ cuprates (κ-Cl: ~$1.8\times10^{14}$ cm$^{-2}$, YBa$_2$Cu$_3$O$_{7-\delta}$: ~$6.7\times10^{14}$ cm$^{-2}$). According to the Hall effect and charge displacement current measurements, the carrier density in κ-Cl can be reversibly tuned up to approximately ±20% (Supplementary Text and Fig. S2). In addition, the effective pressure of the sample can be simultaneously tuned by applying a bending strain to the substrate *(6)*. Therefore, this combination of the EDLT doping and bending-strain techniques allows us to investigate the doping ($\delta$)–correlation ($U/t$)–temperature ($T$) phase diagram for the same sample ($U$ : on-dimer Coulomb repulsion).

## Results

First, we study the resistivity without gating. Figure 1E shows the temperature dependence of the resistivity $\rho$ for various values of the tensile strain $S$ at zero gate voltage. As can be seen, the resistivity of κ-Cl is highly sensitive to the strain because it is in close proximity to the pressure-induced first-order Mott insulator / superconductor transition *(13,14)*. κ-Cl, which is originally a Mott insulator at ambient pressure, is a superconductor on a PET substrate owing to the compressive strain from the substrate induced by its thermal contraction at low temperatures. Upon applying the tensile strain $S$, κ-Cl exhibits a strain-induced superconductor-to-insulator transition at the lowest temperature. At higher temperatures, the resistivity shows a nonmonotonic dependence, changing from semiconducting to metallic with decreasing temperature, followed by resistivity jumps to antiferromagnetic Mott insulating states at lower temperatures. This peculiar dependence on the temperature indicates that the strained κ-Cl is in the proximity of the superconductor / antiferromagnetic Mott insulator transition at half filling, similarly to the bulk sample under hydrostatic pressure *(14-16)*.

Next, we study the resistivity under gating. For $S \leq 0.44\%$, ambipolar superconductivity is observed under gating (Figs. 1D and 2). Upon applying negative gate voltages, the resistivity monotonically decreases by orders of magnitude and superconductivity emerges for $V_G \leq -0.3$ V at $S = 0.41\%$ (Fig. 2C). The critical temperature $T_C$ of approximately 12 K does not vary significantly down to $V_G = -0.5$ V.

On the other hand, the effect of $V_G$ is not monotonic for electron doping. The resistivity abruptly drops and a superconducting state with $T_C$ of approximately 12 K again emerges with low electron doping ($+0.14$ V $\leq V_G \leq +0.22$ V at $S = 0.41\%$). However, the resistivity increases again and the superconductivity disappears after further electron doping. Although the hole- and electron-doped superconducting states show similar values of $T_C$, the doping concentration where superconductivity appears is significantly doping-asymmetric as shown in Figs. 2B-D. Since the charge neutrality point (highest resistivity point) corresponding to half filling is approximately $+0.05$ V in this sample, the electron-doped superconductivity is observed only in a narrow region of $V_G$ near half filling. These superconducting states are suppressed by magnetic fields and their normal states are insulating (Fig. S3), as in the pressure-induced superconducting state at half filling *(17)*. Curiously, the normal-state resistivity takes a local minimum against $V_G$ at the optimum doping on the electron-doped side, as shown in the upper part of Figs. 2B-D. The "dip" of the resistivity is observed at least up to 80 K, which is several times higher than $T_C$ (Fig. S4). The superconducting regions in the $V_G$–$T$ plots become narrower and eventually disappear with increasing tensile strain (Figs. 2F and G). Both the p-type and n-type superconductivity fade at almost the same value of strain.

Summarizing the tensile-strain dependence of $\rho$ in the $V_G$–$T$ plots (Figs. 2A-G), we obtain the $V_G$–$S$ plots at 5.5 K (Fig. 2H), which should reflect the ground-state bandfilling–bandwidth phase diagram. (However, note that the sample conductance is the sum of the doped surface conductance and the nondoped bulk conductance. Namely, the sample resistivity deviates from the doped surface resistivity unless the bulk conductance is much less than the surface conductance. In particular, the resistivity data at $S = 0.35\%$ below 12 K do not reflect the doped states. Once the non-doped bulk becomes superconducting, it cannot be determined whether the doped surface is superconducting or not.)

While the hole-doped superconducting phase lies slightly away from the antiferromagnetic insulating phase, the electron-doped superconducting phase is located in a very narrow region next to the antiferromagnetic insulating phase. The resistive transition between the half-filled insulating state and the electron-doped superconducting state is very abrupt ($\rho$ falls by up to eight orders of magnitude within a gate voltage range of ~0.1 V). At the boundary, discontinuous fluctuations in resistivity are also observed (Fig. S5). These behaviors are reminiscent of the pressure-induced phase-separation/percolative superconductivity at half filling *(14-16)*. Therefore, it is reasonable to consider that the first-order transition line at least extends from half filling to the electron-doped superconducting phase boundary in the diagram.

To summarize the experimental results, [1] the superconducting region is narrower and closer to half filling under electron doping than under hole doping, [2] the electron-doping-driven superconducting transition is very abrupt and discontinuous (first-order-like), [3] the superconducting regions appear to be connected to each other in the $V_G$–$S$ plots within the data resolution, and [4] the doping asymmetry for the resistivity diminishes with decreasing bandwidth.

In contrast to other solid-state materials such as the high-$T_C$ cuprates and doped fullerenes, the bandfilling and bandwidth are simultaneously controlled in a single sample and thus the sample dependence is less significant here. Furthermore, one simple molecular orbital (highest occupied molecular orbital) governs the electronic properties under both electron and hole doping in κ-Cl. Accordingly, the observed doping asymmetry is attributed to the intrinsic electron–hole asymmetry of interacting fermions on the anisotropic triangular lattice.

For better understanding of the particle–hole asymmetry, we employed the variational cluster approximation (VCA) *(18)* to theoretically consider antiferromagnetic and $d_{x^2-y^2}$ superconducting orders in the Hubbard model on the anisotropic triangular lattice defined by the following Hamiltonian:

$$\hat{H} = -\sum_{\langle ij \rangle, \sigma} t_{ij}\left(\hat{c}_{i\sigma}^{\dagger}\hat{c}_{j\sigma} + \text{H.c.}\right) + U\sum_{i}\hat{n}_{i\uparrow}\hat{n}_{i\downarrow} - \mu\sum_{i\sigma}\hat{n}_{i\sigma}, \quad (1)$$

where $\hat{c}_{i\sigma}^{\dagger}$ creates an electron site $i$ with spin $\sigma(=\uparrow, \downarrow)$, $\hat{n}_{i\sigma} = \hat{c}_{i\sigma}^{\dagger}\hat{c}_{j\sigma}$, $t_{ij}$ is the transfer integral between neighboring sites $i$ and $j$ (indicated as $t$ and $t'$ in Fig. 1A), $U$ is the on-site Coulomb repulsion, and $\mu$ is the chemical potential. In the following, we set $t'/t = -0.44$ *(8)* and used a cluster of size $L = 4 \times 3$. We performed scans at 0 K, varying $\mu$ for several values of $U/t$. The strain dependence of the intradimer charge degree of freedom is not considered because the effective Coulomb repulsion on the dimer ($= U$) is not sensitive to changes in the intradimer transfer integral *(5, 19)*: since $U \sim 2|t_{id}| - 4t_{id}^2/U_{bare}$, where $t_{id}$ is the intradimer transfer integral (~0.25 eV) and $U_{bare}$ is the on-site Coulomb repulsion for a single BEDT-TTF molecule (~1.0 eV), small changes in $t_{id}$ cancel out. Here, we employed the one-band model to obtain an approximate outline of the phase diagram. However, note that it is not perfect for describing the electronic properties of κ-type BEDT-TTF salts. Indeed, different gap symmetry (extended-*s* + *d* wave symmetry) has been predicted using the four-band models *(20-22)* and observed by STM in a superconducting salt *(23)*.

Figure 3A shows the antiferromagnetic and superconducting order parameters as a function of doping concentration $\delta = n - 1$, where $n$ is the electron density and $\delta = 0$ corresponds to half filling. Similarly to the experiment, the antiferromagnetic order breaks down and superconductivity emerges at higher $\delta$. Even with a small amount of electron doping ($\delta < 0.1$), the transition takes place when $U/t$ is small. In such a case, the superconductivity appears more abruptly than for hole doping. However, this abrupt breakdown of the antiferromagnetic insulating phase disappears when $U/t$ is increased to more than 5.0.

In addition, we found a nonmonotonic dependence of the chemical potential on $\delta$ under electron doping (Fig. 3B). Accordingly, Fig. 3A includes metastable and unstable solutions on the electron-doped side for small $U/t$ (indicated as empty symbols). As shown in Fig. 3C, the Maxwell construction reveals phase separation between two phases with different doping concentrations $\delta_1$ and $\delta_2$, where the volume fraction of each phase is proportional to $\delta - \delta_1$ or $\delta_2 - \delta$ for the average doping concentration $\delta$, assuming $\delta_1 \leq \delta \leq \delta_2$. If one of the phases is superconducting, percolation superconductivity can appear when $\delta$ exceeds the percolation limit. Although the results depend quantitatively on the clusters used, we found that the tendency towards phase separation is robust under electron doping (see Section S2).

According to the resistivity behavior without a gate voltage (Fig. S6), the change in $U/t$ from $S$ = 0.35 to 0.55% is comparable or slightly larger than that from bulk crystals of κ-(BEDT-TTF)$_2$Cu[N(CN)$_2$]Br (κ-Br, $U/t$ = 5.1, superconducting) to κ-Cl ($U/t$ = 5.5, Mott insulating) *(8)*. Namely the variation in the experiments (~8% change in $t$ for 0.2% increase in strain) is considered to be smaller than that in the calculations shown in Fig. 3A. In addition, the calculations indicate the Mott transition at half filling between $U/t$ = 3.5 and 4.0, while it occurs between κ-Br ($U/t$ = 5.1) and κ-Cl ($U/t$ = 5.5) in the real material system. Despite these discrepancies, the characteristic features in the proximity of the transitions such as the doping asymmetry and the phase separation on the electron-doped side are qualitatively reproduced by our calculations.

The tendency of the phase separation with the sign of doping is opposite to that in the Hubbard model on a square lattice with the model parameters relevant for cuprates, where the phase separation is more likely to occur under hole doping *(24)*. We attribute the phase-separation tendency to the high density of states around the bottom of the upper Hubbard band split off from the van Hove singularity (VHS) by the electron correlation (Fig. 3E). The high density of states accumulated along the Z–M line is characteristic of the anisotropic triangular lattice owing to the fact that the direction of the hopping $t'$ is orthogonal to the Z–M line, along which the dispersion remains flat. The calculations imply that our experiment was able to capture such a highly nontrivial correlated band structure on the unoccupied side as an abrupt change in the transport properties under electron doping.

**Discussion**

The doping and correlation dependence of $T_C$ in the Hubbard model on an anisotropic triangular lattice has also been calculated using cluster dynamical mean-field theory (CDMFT) *(25)*. It predicts hole-doped superconducting states in a wide doping region from extremely low to moderate hole doping (from 1 to 10%), while no superconductivity appears under low electron doping of less than 10%. $T_C$ tends to be enhanced under hole doping and reduced under electron doping, compared with the value at half filling. However, in our experimental results shown in Fig. 2, there are many discrepancies from these theories as well as our previous expectations *(5)*. An abrupt first-order superconducting transition was expected, but it is only observed in the *electron*-doped region. The recurrent insulating phase in the heavily electron-doped region is also a new observation. The values of $T_C$ are unexpectedly independent of both doping and strain. In addition, the electron-doped superconductivity seems to persist at higher $U/t$ than the hole-doped superconductivity. Note that the optimal doping concentration does not change significantly when the strain is varied. These are all new findings observed by the two-dimensional scanning of the strain and carrier density at the κ-Cl surface, although their origins require further discussion.

The disappearance of superconductivity and the appearance of an insulating phase for large electron doping ($V_G \gtrsim +0.25$ V) cannot be explained by the calculations, for example. To understand this, we should probably be aware of the simplicity of the model considered here, which ignores elements specific to this material, such as the intradimer charge degree of freedom and the intersite Coulomb repulsion *(26)*. One possibility is that a magnetic or charge-ordered state emerges at specific doping levels (for example, ~ 12.5%), as in the case of the stripe order in the cuprates *(27-30)*. Indeed, recent theoretical calculations for the Hubbard model on a square lattice suggest that a large part of the macroscopic phase-separation region can be replaced by more microscopically inhomogeneous stripe states *(31,32)*. The presence of a magnetic or charge-ordered state with high resistivity at 12.5% doping may explain the dip in resistivity in the normal state (Fig. S4).

The correlation strength and bandfilling are the most fundamental parameters that determine unconventional superconductivity in correlated electron systems. To construct an experimental correlation–bandfilling phase diagram, however, it has been necessary to combine data from different materials, resulting in unavoidable material and sample dependences. Obtaining the phase diagram in a single sample is a milestone for understanding the pristine mechanism of unconventional superconductivity. This is achieved here by virtue of the high tunability of the organic Mott insulator. In addition, despite the use of a real material, the high controllability and cleanness of our device are in common with those of quantum simulators *(33)* such as cold atoms in an optical lattice. We hope that these results also serve as a useful reference for quantum simulations of the Fermi-Hubbard model beyond the field of materials science.

**Materials and Methods**

Sample preparation and transport measurement

The source, drain, and gate electrodes (18-nm-thick Au) were patterned on a polyethylene terephthalate substrate (Teflex FT7, Teijin DuPont Films Japan Limited) using photolithography. A thin (~100 nm) single crystal of $\kappa$-(BEDT-TTF)$_2$Cu[N(CN)$_2$]Cl (abbreviated to $\kappa$-Cl) was electrochemically synthesized by oxidizing BEDT-TTF (20 mg) dissolved in 50 ml of 1,1,2-trichloroethane (10% v/v ethanol) in the presence of TPP[N(CN)$_2$] (TPP = tetraphenylphosphonium, 200 mg), CuCl (60 mg), and TPP-Cl (100 mg). After applying a current of 8 µA for 20 h, the thin crystal was transferred into 2-propanol with a pipette and guided onto the top of the substrate. A diagonal of the rhombic crystal, which is usually parallel to the crystallographic *a*- or *c*-axis, was aligned parallel to the direction of the current and strain (although it is unknown which axis was the *a*- or *c*-axis). After the substrate was removed from the 2-propanol and dried, the $\kappa$-Cl crystal was shaped into a Hall bar using a pulsed laser beam with a wavelength of 532 nm. The typical dimensions of the Hall bar sample were approximately 15 µm (width) × 40 µm (length) × 100 nm (thickness) as shown in Fig. S1. Figure S1C shows an atomic force microscopy (AFM) image of the surface of a typical $\kappa$-Cl crystal laminated on the substrate. The roughness of the surface was suppressed to less than 1.5 nm over micrometer scale (Fig. S1D), which corresponds to the thickness of one set of BEDT-TTF and anion layers. As a gate electrolyte, the ionic liquid 1-ethyl-3-methylimidazolium 2-(2-methoxyethoxy) ethylsulfate was added dropwise to the sample and a Au side gate electrode. The electric-double-layer transistor (EDLT) device was completed by mounting a 1.2-µm-thick polyethylene naphthalate (PEN) film on the ionic liquid droplet. Thinning of the gate electrolyte using the PEN film reduced the thermal stress at low temperatures.

The transport measurements were performed using a Physical Property Measurement System (Quantum Design). The four-terminal resistance was measured with a dc current of 1 µA using a dc source (Yokogawa 7651, Yokogawa) and a nanovoltmeter (Agilent 34420A, Agilent Technologies). The maximum applied voltage was limited to 1 V (namely, the measurement was voltage-biased in the high-resistance states). The applied current was monitored with a current amplifier (SR570, Stanford Research Systems). The resistivity was estimated from the resistance and sample dimensions. In Fig. 1E, the resistivity $\rho$ ($\Omega$cm) was estimated as $\rho = R \times W \times D/L$ where $R, W, D,$ and $L$ denote the resistance, width, thickness, and length of the sample, respectively. In other figures, the sheet resistivity $\rho = R \times W/L$ is shown because only the sample surface is doped and the three-dimensional resistivity cannot be accurately estimated. According to Hall measurements *(11)*, the effect of doping is confined within one molecular layer under sufficient doping.

The four-terminal resistivity was measured during temperature cycles between 220 and 5 K. The temperature was varied at rates of 2 K/min ($T > 20$ K) and 0.3 K/min (T < 20 K). The gate voltage $V_G$ was swept from +0.5 to −0.5 V with a step of 0.05 V at 220 K, but in the low-electron-doping regime from +0.22 to 0.08 V it was more finely tuned with a step of 0.02 V. After varying the gate voltage, we waited one minute for the stabilization of the gate bias before cooling the sample.

Since the samples are top-gate and bottom-contact transistors, the current flows through the ungated regions between the electrodes and the doped surface, resulting in non-negligible contact resistance. Figures S1E and F show the four-terminal and two-terminal resistances without and with a gate voltage, respectively. One can see that the temperature dependence of the contact resistance is more moderate than that of the four-terminal resistance. Therefore, the ratio of contact resistance to sample resistance is large when the sample is metallic or superconducting. However, the contact resistance is typically up to 10 k$\Omega$ order and does not hinder the four-terminal resistivity measurement.

After the temperature cycles at different gate voltages, a tunable tensile strain was mechanically applied with a nanopositioner (ANPz51, attocube) from the back side of the substrate at 220 K. The basic techniques and apparatus for the strain measurements were the same as those in our previous report *(6)*. The tensile strain was estimated from $S = 4tx/(l^2 + 4x^2)$, where $t$ and $l$ are the thickness and length of the substrate, respectively, and $x$ is the displacement of the nanopositioner. We assume that [1] the bent substrate (and κ-Cl crystal) is an arc of a circle, [2] the angle between the ends of the substrate is small (small angle approximation). The tensile strain was applied as the following order: 0.39, 0.41, 0.44, 0.46, 0.50, 0.55, and 0.35%.

VCA calculations

1. Model Hamiltonian

In order to study the ground-state phase diagram of the organic Mott insulator κ-Cl from the theoretical point of view, we consider the single-band Hubbard model on the anisotropic triangular lattice *(8,33-36)*. The Hamiltonian of the model is

$$\hat{H} = -\sum_{\langle ij \rangle, \sigma} t_{ij} \left( \hat{c}^\dagger_{i\sigma} \hat{c}_{j\sigma} + \text{H.c.} \right) + U \sum_i \hat{n}_{i\uparrow} \hat{n}_{i\downarrow} - \mu \sum_{i\sigma} \hat{n}_{i\sigma}, \qquad (1)$$

where $\hat{c}^\dagger_{i\sigma}$ creates an electron site $i$ with spin $\sigma(=\uparrow,\downarrow)$, $\hat{n}_{i\sigma} = \hat{c}^\dagger_{i\sigma} \hat{c}_{i\sigma}$, $t_{ij}$ is the transfer integral between neighboring sites $i$ and $j$ (indicated as $t$ and $t'$ in Fig. 1A), $U$ is the on-site Coulomb repulsion, and $\mu$ is the chemical potential. The transfer integral between the different (same) dimers of BEDT-TTF molecules is given by $t_{ij} = t$ ($t'$). We set $t'/t = -0.44$, which is relevant for κ-Cl *(8)*, and vary $U$ and $\mu$ to control the strength of the electron correlation and the carrier concentration, respectively. In experiment, $U/t$ and $\mu$ can be controlled by applying the strain $S$ and the gate voltage $V_G$, respectively. Namely, the tensile strain $S$ increases $U/t$, and the positive (negative) gate voltage increases (decreases) $\mu$.

2. Variational cluster approximation

We apply the variational cluster approximation (VCA) *(37)* to the single-band Hubbard model in Eq. (1). The VCA is a many-body variational method based on the self-energy-functional theory (SFT) *(12,38,39)*, and allows us to investigate the possible spontaneous symmetry breaking including antiferromagnetism (AFM) and superconductivity (SC) by taking into account the short-range electron correlations through the exact self energy of a so-called reference system which will be introduced later. The VCA has been applied for the single-band Hubbard model on the anisotropic triangular lattice at half filling *(40-42)* and also on the square lattice with finite dopings relevant for cuprates *(24,43)* to study possible magnetism and superconductivity. Moreover, the VCA has been able to capture the first-order phase transtions of strongly correlated systems, as well as the free-energy balance and the Maxwell construction consistently *(44-46)*.

Let us first review the variational principle for the grand potential on which the VCA is constructed *(39,47,48)*. There exists a functional $\Omega[\Sigma]$ of the self energy $\Sigma$ that gives the grand potential $\Omega$ as $\Omega = \Omega[\Sigma^*]$ such that

$$\left. \frac{\delta \Omega[\Sigma]}{\delta \Sigma} \right|_{\Sigma = \Sigma^*} = 0, \qquad (2)$$

where

$$\Omega[\Sigma] = F[\Sigma] - \frac{1}{\beta}\text{Tr}\ln(-\boldsymbol{G}_0^{-1} + \Sigma), \qquad (3)$$

$\boldsymbol{G}_0$ is the noninteracting single-particle Green's function, $F[\Sigma]$ is the Legendre transform of the Luttinger-Ward functional $\Phi[\boldsymbol{G}]$ *(45)*, i.e., $F[\Sigma] = \Phi[\boldsymbol{G}] - \beta^{-1}\text{Tr}[\boldsymbol{G}\Sigma]$ with $\Sigma = \beta\delta\Phi[\boldsymbol{G}]/\delta\boldsymbol{G}$, $\boldsymbol{G}$ is the interacting single-particle Green's function, $\beta = 1/T$ is the inverse temperature, and Tr denotes the functional trace that runs over the all (both spatial and temporal, regardless of discrete or continuous) variables of the functions in Tr[$\cdot \cdot \cdot$]. The stationary condition Eq. (2) ensures that $\Sigma^*$ is the physical self energy in the sense that $\Sigma^*$ satisfies the Dyson equation. The variational principle for the grand potential has been derived by the diagrammatic-expansion technique for many-body Green's function *(47)*, and later by the functional-integral technique for the grand-partition function in a nonperturbative way *(48)*. A remarkable property of the Luttinger-Ward functional is that its functional form F depends only on the interaction term of the Hamiltonian *(48)*.

Although the variational principle is rigorous and nonperturbative, a difficulty in applying it for practical calculations is that the explicit form of the Luttinger-Ward functional $F[\Sigma]$ is unknown. The key approximation of the VCA is to restrict the space of the trial self energy $\Sigma$ to that of the exact self energy $\Sigma'$ of a reference system described by the Hamiltonian $\widehat{H}'$ whose interaction term must remain unchanged but the single-particle terms can be modified from $\widehat{H}$ in order for $\widehat{H}'$ to be solvable exactly. The invariance of the Luttinger-Ward functional $F$ enables us to write the grand-potential functional of reference system as $\Phi'[\Sigma'] = F[\Sigma] - \beta^{-1}\text{Tr}\ln(-\boldsymbol{G}_0'^{-1} + \Sigma')$, where $\boldsymbol{G}'_0$ is the noninteracting Green's function of the reference system. By restricting the trial self energy to $\Sigma'$ and eliminating $F[\Sigma]$ from Eq. (3), we obtain

$$\Omega[\Sigma'] = \Omega'[\Sigma'] - \frac{1}{\beta}\text{Tr}\ln(\boldsymbol{I} - \boldsymbol{V}\boldsymbol{G}'[\Sigma']), \qquad (4)$$

where $\boldsymbol{V} = \boldsymbol{G}'^{-1}_0 - \boldsymbol{G}_0^{-1}$ represents the difference of the single-particle terms between the original and reference systems and $\boldsymbol{G}'[\Sigma'] = (\boldsymbol{G}'^{-1}_0 - \Sigma')^{-1}$ is the exact interacting single-particle Green's function of the reference system. Note that the right-hand side of Eq. (4) is computable as long as the reference system $\widehat{H}'$ is solvable.

The remaining arbitrariness of the single-particle terms in the reference system $\widehat{H}'$ allows us to optimize the trial self energy $\Sigma'$ through varying single-particle fields $\lambda$, which parametrize the single-particle terms, as variational parameters, i.e., $\Sigma' = \Sigma'(\lambda)$, so as to satisfy the stationary condition in Eq. (2). Simply denoting $\Omega(\lambda) = \Omega[\Sigma'(\lambda)]$, the optimization scheme of the VCA amounts to find the extrema

$$\left.\frac{\partial \Omega(\lambda)}{\partial \lambda}\right|_{\lambda=\lambda^*} = 0, \quad (5)$$

in the space of the variational parameters $\lambda$, where $\lambda^*$ denotes the set of optimal variational parameters.

3. Reference system

Now we specify our reference system $\hat{H}'$ used for the VCA calculations. We assume that our reference system is composed of a collection of identical, disconnected, and finite-size clusters, each of which is described by a Hamiltonian $\hat{H}_c(\boldsymbol{R}_{i_c})$, i.e., $\hat{H}' = \sum_{i_c=1}^{N} \hat{H}_c(\boldsymbol{R}_{i_c})$, where $\boldsymbol{R}_{i_c}$ is the $i_c$-th position of the cluster and $N$ is the number of the clusters in the reference system. Since the clusters are identical, i.e., $\hat{H}_c(\boldsymbol{R}_{i_c}) = \hat{H}_c$, we refer to it as $\hat{H}_c$.

In order to study SC and AFM phases under the carrier doping, we consider the following Hamiltonian of the cluster:

$$\hat{H}_c = \hat{H}_H + \hat{H}_\epsilon + \hat{H}_h + \hat{H}_\Delta, \quad (6)$$

where

$$\hat{H}_\epsilon = \epsilon \sum_i (\hat{n}_{i\uparrow} + \hat{n}_{i\downarrow}), \quad (7)$$

$$\hat{H}_h = h \sum_i (\hat{n}_{i\uparrow} - \hat{n}_{i\downarrow}) e^{i\boldsymbol{Q}\cdot\boldsymbol{r}_i}, \quad (8)$$

$$\hat{H}_\Delta = \Delta \sum_{\langle i,j \rangle} (\eta_{ij} \hat{c}_{i\downarrow} \hat{c}_{j\uparrow} + \text{H.c.}), \quad (9)$$

and $\hat{H}_H$ is the Hubbard Hamiltonian given in the right-hand side of Eq. (1) but defined in the cluster. Here, $\boldsymbol{Q} = (\pi, \pi)$ and $\boldsymbol{r}_i$ is the position of the $i$-th site. $\epsilon$ is the variational parameter for the on-site potential that has to be optimized in order to calculate the particle density satisfying the thermodynamic consistency (24). $\Delta$ and $h$ are the variational parameters that are introduced to detect the $d$-wave SC and AFM states by explicitly breaking the corresponding U(1) and SU(2) symmetries in the reference system, respectively. The spatial dependence of the form factor $\eta_{ij}$ for the superconductivity considered here is given as

$$\eta_{ij} = \begin{cases} +1 & \text{if } \boldsymbol{r}_i - \boldsymbol{r}_j = \pm \boldsymbol{e}_1, \\ -1 & \text{if } \boldsymbol{r}_i - \boldsymbol{r}_j = \pm \boldsymbol{e}_2, \\ 0 & \text{otherwise}, \end{cases} \quad (10)$$

where $\mathbf{e}_1$ and $\mathbf{e}_2$ are the primitive lattice vectors on the anisotropic triangular lattice bridging the two sites connected by the nearest-neighbor hopping integral $t$ (not by $t'$). Note that the form factor $\eta_{ij}$ corresponds to that of the $d_{x^2-y^2}$-wave SC if it is considered on the square lattice spanned by the primitive lattice vectors $\mathbf{e}_1$ and $\mathbf{e}_2$ (assuming that $\mathbf{e}_1$ and $\mathbf{e}_2$ point to the $x$ and $y$ directions, respectively). Therefore, we refer to the SC state found in the VCA as $d_{x^2-y^2}$-SC.

4. Particle-hole transformation

Since the pairing term $\widehat{H}_\Delta$ in Eq. (9) explicitly breaks the U(1) symmetry, the particle number is not conserved as long as $\Delta$ is finite. This is inconvenient if the computer program of the exact-diagonalization method is implemented with the number of particles fixed. However, since the $z$ component of the total spin is conserved, it is still possible to work in the basis set for the fixed number of particles after the particle-hole transformation

$$\begin{cases} \hat{c}_i := \hat{c}_{i\uparrow} \\ \hat{d}_i^\dagger := \hat{c}_{i\downarrow}. \end{cases} \quad (11)$$

In terms of the newly defined spinless-fermion operators $\hat{c}_i$ and $\hat{d}_i$, the single-particle terms in Eqs. (7)–(9) become

$$\widehat{H}_\epsilon = \epsilon \sum_i (\hat{n}_{ic} - \hat{n}_{id} + 1), \quad (12)$$

$$\widehat{H}_h = h \sum_i (\hat{n}_{ic} + \hat{n}_{id} - 1) e^{i\mathbf{Q}\cdot\mathbf{r}_i}, \quad (13)$$

$$\widehat{H}_\Delta = \Delta \sum_{\langle i,j \rangle} (\eta_{ij} \hat{d}_i^\dagger \hat{c}_j + \text{H.c.}), \quad (14)$$

where $\hat{n}_{ic} = \hat{c}_i^\dagger \hat{c}_i$ and $\hat{n}_{id} = \hat{d}_i^\dagger \hat{d}_i$. Note that $\widehat{H}_\Delta$ becomes the hybridization between $c$ and $d$ orbitals. By applying the particle-hole transformation also for $\widehat{H}_H$, the Hamiltonian of the cluster is now given as

$$\widehat{H}_c = -\sum_{\langle ij \rangle} t_{ij} (\hat{c}_i^\dagger \hat{c}_j + \text{H.c.}) + \sum_{\langle ij \rangle} t_{ij} (\hat{d}_i^\dagger \hat{d}_j + \text{H.c.})$$

$$+ (U - \mu + \epsilon + h) \sum_i \hat{n}_{ic} + (\mu - \epsilon - h) \sum_i \hat{n}_{id} \quad (15)$$

$$- U \sum_i \hat{n}_{ic} \hat{n}_{id} + \Delta \sum_{\langle ij \rangle} (\eta_{ij} \hat{d}_i^\dagger \hat{c}_j + \text{H.c.}) - \mu L + \epsilon L - h \sum_i e^{i\mathbf{Q}\cdot\mathbf{r}_i},$$

where L is the number of sites in a cluster. It is now apparent that the total number of the particles is conserved as the operator commutes with the Hamiltonian $\hat{H}_c$, which is inherited from the conservation of the z component of the total spin before the particle-hole transformation.

Here, several remarks on the non-operator terms (i.e., constant terms) in Eq. (15) are in order. Unlike the single-particle operator terms multiplied by the variational parameters, the non-operator terms in the last line of Eq. (15) cannot be "subtracted" as perturbation in the $V$ matrix in Eq. (4), since the $V$ matrix is a representation of the single-particle operators *(35)*. Instead, those numbers have to be subtracted either from $\hat{H}_c$ itself or from the eigenspectrum of the cluster Hamiltonian $\hat{H}_c$ [see for example Eq. (16)]. Note also that the term $-\mu L$ in the third line of Eq. (15) has to be kept because μ is the model parameter that has the definite physical meaning. Finally, we comment on the number of the basis states in the cluster. Since we consider the ground state of the cluster at zero temperature, we can focus on the sector for which the z component of the total spin is zero, i.e., $\sum_{i=1}^{L}(\hat{n}_{i\uparrow}-\hat{n}_{i\downarrow})|x\rangle = 0|x\rangle$, where $|x\rangle$ is an arbitrary basis state for the corresponding sector. After the particle-hole transformation of Eq. (11), this sector corresponds to that with the total number of particles being L, i.e., $\sum_{i=1}^{L}(\hat{n}_{ic}+\hat{n}_{id})|x\rangle = L|x\rangle$. Since the cluster consists of $2L$ orbitals with $L$ particles, the number of configurations in this sector is given by the binomial coefficient $\binom{2L}{L}$. In the main text, we consider the cluster of the 4 × 3 sites ($L = 12$), for which the number of the configuration is $\binom{24}{12} = 24!/(12!)^2 = 2704156$. Considering that the VCA has to calculate the ground state and the single-particle Green's function of the cluster repeatedly until the stationary condition Eq. (5) is satisfied, and also that the model parameters $U/t$ and $\mu/t$ are swept finely in a wide range in order to find not only stable but also metastable and unstable solutions that can identify the regions of the phase separation or the first-order transitions, $L = 12$ is reasonably large among the clusters feasibly handled by the exact-diagonalization method.

5. Grand-potential functional and order parameters

Having specified the reference system, we can further substantiate the grand-potential functional. Since the system is considered at equilibrium and the clusters in the reference system are periodically aligned, the summand of the functional trace Tr[· · ·] in Eq. (4) is diagonal with respect to the Matsubara frequencies and the wavevectors of the superlattice on which the clusters are lined up. The grand-potential functional per site can hence be written as

$$\Omega = \Omega' - \frac{1}{NL\beta} \sum_v \sum_{\tilde{k}} \ln \det[I - V(\tilde{k})G'(i\omega_v)], \quad (16)$$

where

$$\Omega' = -\frac{1}{L\beta} \ln \sum_s e^{-\beta(E_s - \tilde{E})}, \quad (17)$$

$E_s$ is the eigenvalue of $\hat{H}_c$ for $s$-th eigenstate $|\Psi_s\rangle$ with the ground state $|\Psi_0\rangle$ (note that $\hat{H}_c$ includes the chemical potential term in our convention), $\tilde{E} = \epsilon L - h \sum_i e^{i\mathbf{Q}\cdot\mathbf{r}_i}$ is the non-operator term in $\hat{H}_c$ as annotated after Eq. (15), $\omega_v = (2v+1)\pi/\beta$ is the fermionic Matsubara frequency with $v$ integer, and $\tilde{k}$ is the wavevector defined in the Brillouin zone of the superlattice with the number of wavevectors $\tilde{k}$ being $N$. Here we have re-defined $\Omega$ and $\Omega'$ as the corresponding quantities $\Omega[\Sigma]'$ and $\Omega'[\Sigma']$ per site. $G'(z)$ represents the exact single-particle Green's function of the cluster, which is given as

$$G'^{cd}_{ij}(z) = \sum_s e^{\beta(\Omega' - E_s)} \left( G^{+,cd}_{ij,s}(z) + G^{-,cd}_{ij,s}(z) \right), \quad (18)$$

where

$$G^{+,cd}_{ij,s}(z) = \langle \Psi_s | \hat{c}_i [z - (\hat{H}_c - E_s)]^{-1} \hat{d}_j^\dagger | \Psi_s \rangle, \quad (19)$$

$$G^{-,cd}_{ij,s}(z) = \langle \Psi_s | \hat{d}_j^\dagger [z + (\hat{H}_c - E_s)]^{-1} \hat{c}_i | \Psi_s \rangle, \quad (20)$$

and similarly for $G'^{cc}_{ij}(z)$, $G'^{dc}_{ij}(z)$, and $G'^{dd}_{ij}(z)$. The block Lanczos method is adapted for the calculation of $G'(z)$. $V$ represents the hopping process between the clusters and the subtraction of single-particle terms added to the reference system (i.e., $\hat{H}_\epsilon$, $\hat{H}_h$, and $\hat{H}_\Delta$ after the particle-hole transformation without the non-operator term).

For a given set of the model parameters $U/t$ and $\mu/t$, all the variational parameters $\lambda = (\epsilon, h, \Delta)$ are optimized simultaneously in order to satisfy the stationary condition

$$\left( \frac{\partial \Omega(\lambda)}{\partial \epsilon}, \frac{\partial \Omega(\lambda)}{\partial h}, \frac{\partial \Omega(\lambda)}{\partial \Delta} \right) \bigg|_{\lambda = \lambda^*} = (0,0,0), \quad (21)$$

where $\lambda^* = (\epsilon^*, h^*, \Delta^*)$ are the optimal variational parameters. The Newton-Raphson method is used for the optimization. The first and the second derivatives of $\Omega$ with respect to the variational parameters necessary for the gradient and the Hessian matrix are calculated by the central finite-difference method with the error in the second order of the step size. The step size for the finite difference is chosen adaptively by monitoring the curvature of $\Omega$ in the variational-parameter space according to the scheme proposed in Ref. *(45)*.

After the optimal variational parameters $\boldsymbol{\lambda}^*$ are found, the expectation value of a single-particle operator can be evaluated from the VCA Green's function

$$\widetilde{G}(\check{\mathbf{k}}, i\omega_\nu) = G'(i\omega_\nu)\big[I - V(\check{\mathbf{k}})G'(i\omega_\nu)\big]^{-1}\Big|_{\lambda=\lambda^*}. \quad (22)$$

For example, the doping concentration $\delta$, the staggered magnetization $M$, and the $d_{x^2-y^2}$-SC order parameter $D$ is calculated respectively as

$$\delta = \frac{1}{NL\beta} \sum_\nu \sum_{\check{\mathbf{k}}} \mathrm{Tr}\big[\widetilde{G}(\check{\mathbf{k}}, i\omega_\nu)\big], \quad (23)$$

$$M = \frac{1}{2NL\beta} \sum_\nu \sum_{\check{\mathbf{k}}} \mathrm{Tr}\big[\boldsymbol{m}\widetilde{G}(\check{\mathbf{k}}, i\omega_\nu)\big], \quad (24)$$

$$D = \frac{1}{2NB\beta} \sum_\nu \sum_{\check{\mathbf{k}}} \mathrm{Tr}\big[\boldsymbol{\eta}\widetilde{G}(\check{\mathbf{k}}, i\omega_\nu)\big], \quad (25)$$

where $B$ is the number of pairs of sites connected by the $d_{x^2-y^2}$-wave pairing field in the cluster, and $\boldsymbol{m}$ and $\boldsymbol{\eta}$ are real-symmetric $2L \times 2L$ matrices whose matrix elements are either −1, 0, or +1 to represent the staggered magnetization and the $d_{x^2-y^2}$ pairing, respectively. Note that $\widetilde{G}(\check{\mathbf{k}}, i\omega_\nu)$ is assumed to be in the Nambu-spinor representation and thus the trace of $\widetilde{G}(\check{\mathbf{k}}, i\omega_\nu)$ is not the electron density but the doping concentration.

In the zero-temperature limit, the sum $\Sigma_s$ over the eigenstates of $\widehat{H}_c$ in Eqs. (17) and (18) is taken only for the ground state of the cluster ($s = 0$), and the sum $\Sigma_\nu$ over Matsubara frequencies in Eqs. (18) and (23)–(25) is converted to the integral over the continuous frequency along the imaginary axis with a proper regularization for the integrand *(49)*. The number $N$ of wavevectors $\check{\mathbf{k}}$ in the Brillouin zone of the superlattice is chosen adaptively by monitoring the convergence of the sum $\Sigma_{\check{\mathbf{k}}}$ over $\check{\mathbf{k}}$ with respect to $N$ for each frequency. In general, the larger $N$ is required for frequencies closer to the Fermi level to reach the convergence.

**Acknowledgments**

We would like to acknowledge Teijin DuPont Films Japan Limited for providing the PET films. **Funding:** This work was supported by MEXT and JSPS KAKENHI (Grant Nos. JP16H06346, JP15K17714, JP26102012 and JP25000003), JST ERATO, MEXT Nanotechnology Platform Program (Molecule and Material Synthesis), and MEXT HPCI Strategic Programs for Innovative Research (SPIRE) (Grant Nos. hp140128 and hp150112). We are also grateful for allocating computational time of the HOKUSAI GreatWave and HOKUSAI BigWaterfall supercomputer at RIKEN Advanced Center for Computing and Communication (ACCC), and the K computer at RIKEN Center for Computational Science (R-CCS). K.S. acknowledges support from the Overseas Research Fellowship Program of the Japan Society for the Promotion of Science.
**Author contributions:** Y.K. and K.S. contributed equally to this work. Y.K. performed the planning, sample fabrication, cryogenic transport measurements and data analyses. K.S. and S.Y. performed all VCA calculations and data analyses. J.P. and T.T. developed the techniques for ion liquid and improved the device performance. S.T. performed the displacement current measurements and the data analyses. Y.K., K.S. and H.M.Y. wrote the manuscript. T.T., S.Y., H.M.Y. and R.K. supervised the investigation. All authors commented on the manuscript.
**Competing interests:** Authors declare no competing interests. **Data and materials availability:** The data that support the findings of this study are available from the corresponding authors upon request.

**Figures and Tables**

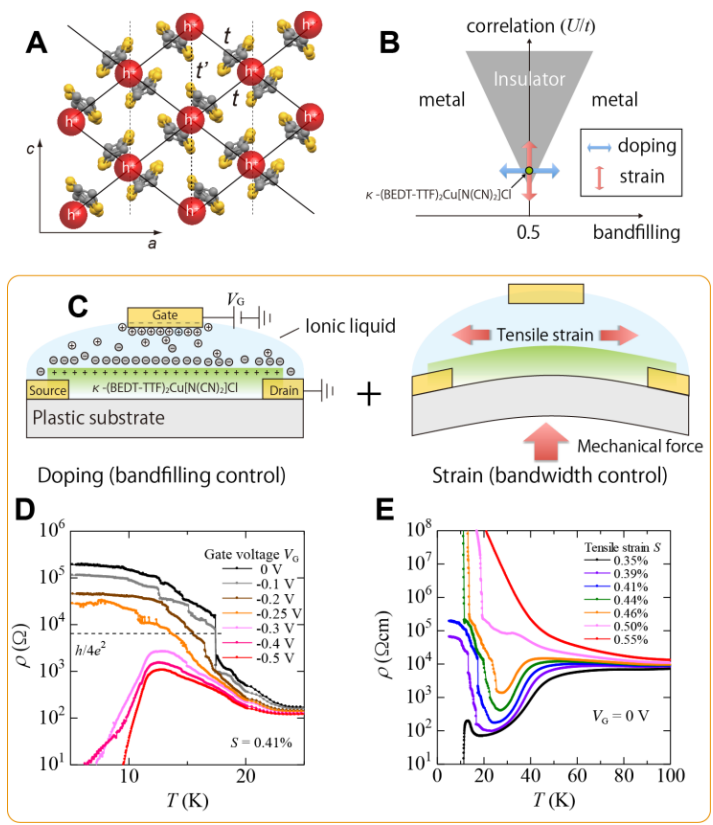

**Fig. 1. Bandfilling and bandwidth control of κ-Cl in the same sample.** (**A**) Molecular arrangement of the BEDT-TTF layer in κ-Cl (top view). (**B**) Conceptual phase diagram based on the Hubbard model *(12)*. The vertical axis denotes the strength of the electron correlation. κ-Cl is originally located near the tip of the insulating region and is shifted along both directions to investigate the superconducting region. (**C**) Schematic side view of the device structure. The doping concentration and effective pressure are controlled by an electric-double-layer gating and bending of the substrate with a piezo nanopositioner, respectively. The resistivity is measured by the standard four-probe method. (**D**) Sheet resistivity vs temperature plots under hole doping at tensile strain $S = 0.41\%$. The dashed line indicates the pair quantum resistance $h/4e^2$. (**E**) Resistivity vs temperature plots under different tensile strains at gate voltage $V_G = 0$ V.

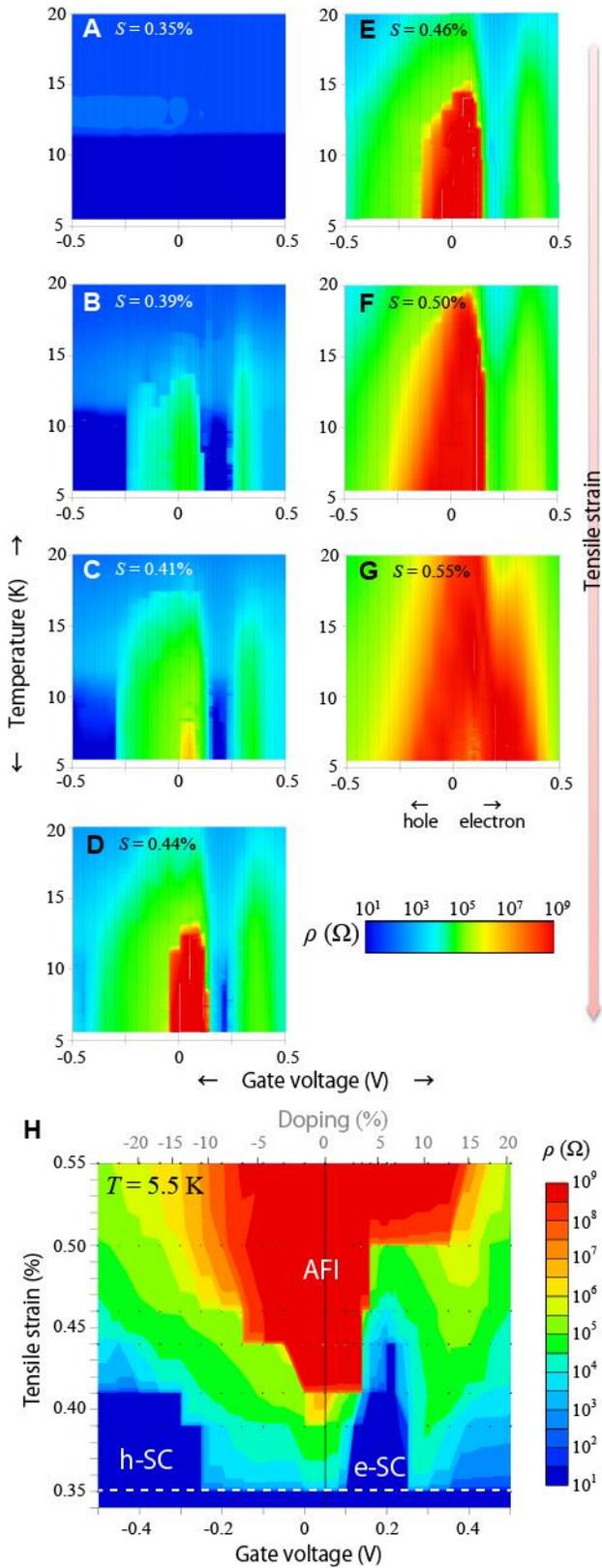

**Fig. 2. Electron–hole asymmetric phase diagram of κ-Cl.** (**A-G**) Contour plots of the sheet resistivity $\rho$ under tensile strains $S = 0.35\%$ (**A**), $0.39\%$ (**B**), $0.41\%$ (**C**), $0.44\%$ (**D**), $0.46\%$ (**E**), $0.50\%$ (**F**), and $0.55\%$ (**G**) as a function of temperature and gate voltage. (**H**) Contour plots of the sheet resistivity ρ at 5.5 K as a function of gate voltage and tensile strain. Black dots in all figures

indicate the data points where the sheet resistivity was measured. The doping concentration estimated from the average density of charge accumulated in the charge displacement current measurement (Fig. S2) is shown for reference on the upper horizontal axis in **H**. AFI, h-SC, and e-SC denote an antiferromagnetic insulator, p-type superconductivity, and n-type superconductivity, respectively. In the region below the white dashed line at $S = 0.35\%$, the surface resistivity under doping cannot be measured because the non-doped bulk is superconducting below 12 K.

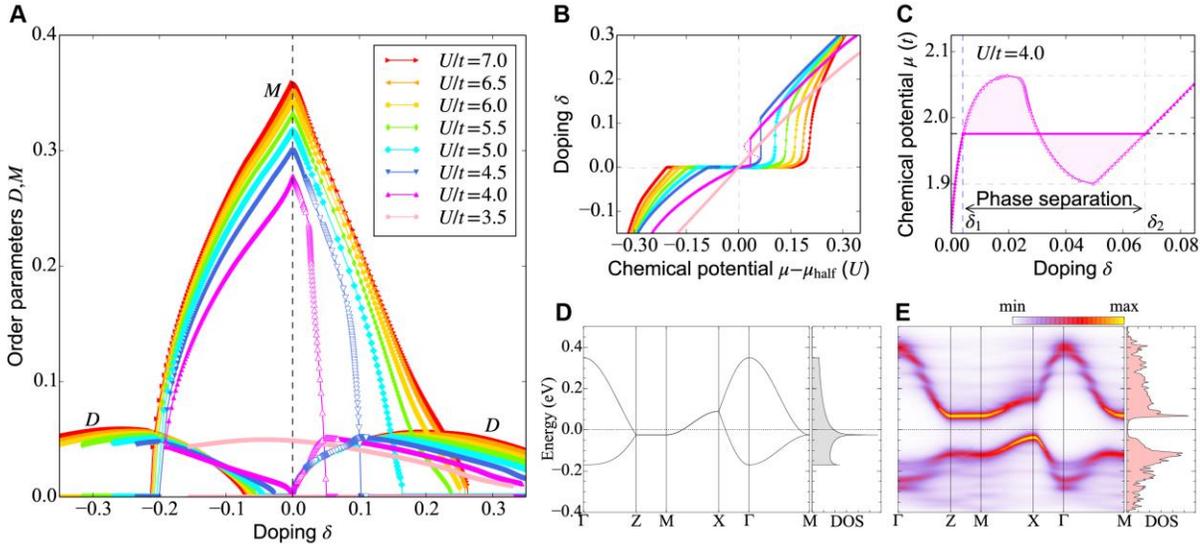

**Fig. 3. VCA calculations.** (**A**) Antiferromagnetic and $d_{x^2-y^2}$ superconducting order parameters, $M$ and $D$, respectively, vs doping concentration $\delta$ for several values of $U/t$. $M$ and $D$ for the metastable and unstable solutions (empty symbols) are also shown at $U/t = 4$ and 4.5 under electron doping (corresponding to positive $\delta$). (**B**) Doping concentration $\delta$ vs chemical potential $\mu$ relative to that at half filling ($\mu_{\text{half}}$) for several values of $U/t$. The results for the metastable and unstable solutions at $U/t = 4$ and 4.5 are indicated by dashed lines, while the results obtained by the Maxwell construction are denoted by solid vertical lines. This implies the presence of phase separation and a first-order phase transition. It is noteworthy that there is a steep (nearly vertical) increase in $\delta$ with increasing $\mu$ for larger values of $U/t$ under electron doping, suggesting a strong tendency towards phase separation. The values of $\mu_{\text{half}}$ are $\mu_{\text{half}} = 1.3725t$ for $U/t = 3.5$, $\mu_{\text{half}} = 1.8375t$ for $U/t = 4$, and $\mu_{\text{half}} = U/2$ for $U/t$ 4.5. (**C**) Chemical potential $\mu$ vs doping concentration $\delta$ for $U/t = 4$ (see Fig. S8 for more details). $\delta_1$ and $\delta_2$ are the doping concentrations of the two extreme states in the phase separation. All results in **A**-**C** are calculated using the VCA for the single-band Hubbard model on an anisotropic triangular lattice ($t'/t = -0.44$) with a $4 \times 3$ cluster.

(**D**) Non-interacting tight-binding band structure and density of states (DOS) for $t'/t = -0.44$ with $t = 65$ meV. Here, $\Gamma = (0,0)$, $Z = (0,\pi/c)$, $M = (\pi/a,\pi/c)$, and $X = (\pi/a,0)$ with $a$ and $c$ being the lengths of the primitive translation vectors indicated in Fig. 1A. The Fermi level for half filling is set to zero and denoted by dashed lines. (**E**) Single-particle spectral functions and DOS of κ-Cl at half filling in an antiferromagnetic state at zero temperature, calculated by variational cluster approximation. The Fermi level is denoted by a dashed line at zero energy. The flat features seen away from the Fermi level indicate incoherent continuous spectra due to the electron correlation. The reason why they appear rather discretized is because of the discrete many-body energy levels in the VCA calculation, for which a finite-size cluster is used to obtain the single-particle excitation energies.

**Supplementary Materials**

**Section S1. Estimation of injected charge density**

In the main text, we controlled the gate voltage $V_G$ to tune the doping concentration of the sample. Although it is not easy to accurately determine the carrier density in strongly correlated electron systems, we roughly estimated the relationship between $V_G$ and the doping concentration by the Hall effect and displacement current measurements. For these measurements, we employed PEN substrates on which κ-Cl behaves as a Mott insulator without applying strain.

The Hall effect determines the sign and density of the carriers flowing in a conductor. In the κ-Cl EDLT, the Mott transition occurs and a large holelike Fermi surface appears under sufficient electron doping *(11)*. Therefore, the injected electron density can be speculated from $n_{HF} - 1/eR_H$, where $n_{HF}$ is the half-filled (non-interacting) hole density of κ-Cl and $R_H$ is the Hall coefficient. According to the Hall effect at 40 K (Fig. S2A), $n_{HF} - 1/eR_H = 3.6\times10^{13} \text{cm}^{-2} \sim 22\%$ at $V_G - V_{CNP} = 0.485$ V, where $V_{CNP}$ is the gate voltage (charge neutrality point) at which the resistivity is highest with varying $V_G$. However, it is difficult to estimate the injected hole density from $R_H$ due to the pseudogap that appears under hole doping.

On the other hand, the displacement current measurement provides information on the accumulated charge density on the sample surface *(50)*. The displacement current $I_{disp}$ is equivalent to the gate current in an ideal sample without any parasitic capacitance, and the accumulated charge density $p$ can be estimated as

$$p = \frac{Q}{eA} = \frac{\int I_{disp} V_G}{r_V eA}, \quad (26)$$

where $Q$ is the total accumulated charge, $A$ is the sample area, and $r_V$ is the sweep rate of $V_G$. Figure S2B shows the $p$ vs $V_G$ plots at different sweep rates. According to these plots, the doping concentration estimated from $p$ is $22.7 \pm 0.8\%$ at $V_G = +0.5$ V (electrons) and $22.5 \pm 0.6\%$ at $V_G = -0.5$ V (holes). Note that these measurements contain errors due to the effects of thermally excited carriers in the bulk (for the Hall coefficient measurement) and a finite parasitic capacitance related to the device structure (for the displacement current measurement). Despite these complications, these two measurements are consistent with each other on the electron-doped side. Note also that the displacement current measurement indicates that the carrier density does not depend strongly on the polarity of $V_G$.

**Section S2. Phase separation tendency of the VCA calculations**

Here, we discuss the phase-separation tendency within the VCA. First, we should note that the results of the VCA calculations generally depend quantitatively on the size of clusters used. A systematic way of treating the finite-size effect by a finite-size scaling has been reported for a Mott insulating state *(49)*. However, such a finite-size scaling is not always satisfactory when the system is in a moderate- or weak-coupling regime or away from half filling. In these cases, the finite-size effect is in general more severe than that in a Mott insulating state because the electrons are more delocalized. Moreover, the number of sites that can be treated by the exact diagonalization method is severely limited due to the exponential increase of the dimension of the Hilbert space. Therefore, it is difficult to perform systematically a finite-size scaling to determine the region of the phase separation in the thermodynamic limit.

Nevertheless, the VCA calculations for several finite-size clusters of $L = 2 \times 2$, $4 \times 2$, and $4 \times 3$ sites find a clear tendency toward the phase separation in the Hubbard model on the anisotropic triangular lattice with $t'/t = -0.44$ (Fig. S7). The features that are observed irrespectively of the clusters are summarized as follows [see also Fig. S8]: (i) The phase separation is not found under hole doping, (ii) the phase separation is not found under electron doping for relatively large $U/t \gtrsim 7$, (iii) the phase separation can be found at small electron-doping concentrations for moderate $U/t$, which is close to the critical $U/t$ of the transition between the superconductivity and the antiferromagnetic insulator at half filling, and (iv) even when the phase separation is absent, the compressibility $\kappa = \partial\delta/\partial\mu$ tends to become much larger under electron doping than under low hole doping, suggesting a strong tendency towards the phase separation under electron doping. These features are qualitatively consistent with the experimental results. Finally, we note that the strong tendency toward the phase separation under electron doping found here is opposite to the case of the Hubbard model on the square lattice with model parameters relevant for cuprates *(24)*, because there the phase separation is more likely under hole doping.

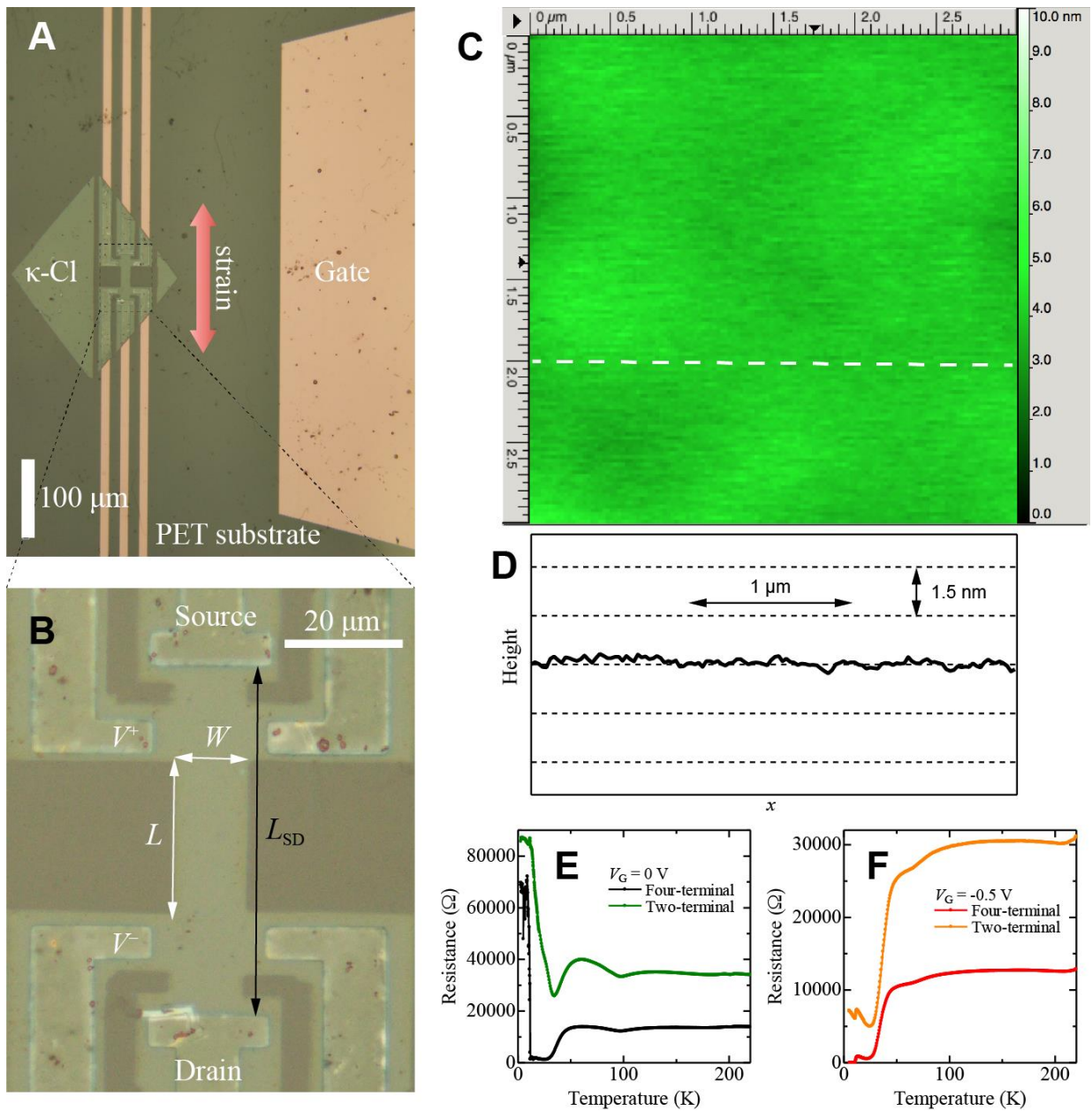

**Fig. S1.**

**Optical images, surface profile, and contact resistance properties of the flexible electric-double-layer transistor based on κ-Cl.** (**A,B**) Optical images of a sample. In the four-terminal resistivity measurement, dc current is applied between the source and drain electrodes, and the voltage difference between the $V^+$ and $V^-$ electrodes is measured. (**C**) Atomic force microscopy (AFM) image and (**D**) line profile of typical κ-Cl crystal laminated on a substrate along the dashed line in **C**. (**E,F**) Four-terminal resistivity (between $V^+$ and $V^-$) and two-terminal resistance (between source and drain) of a sample without (**E**) and with (**F**) a gate voltage. $L_{SD}/L$ (indicated in **B**) is approximately 2.2.

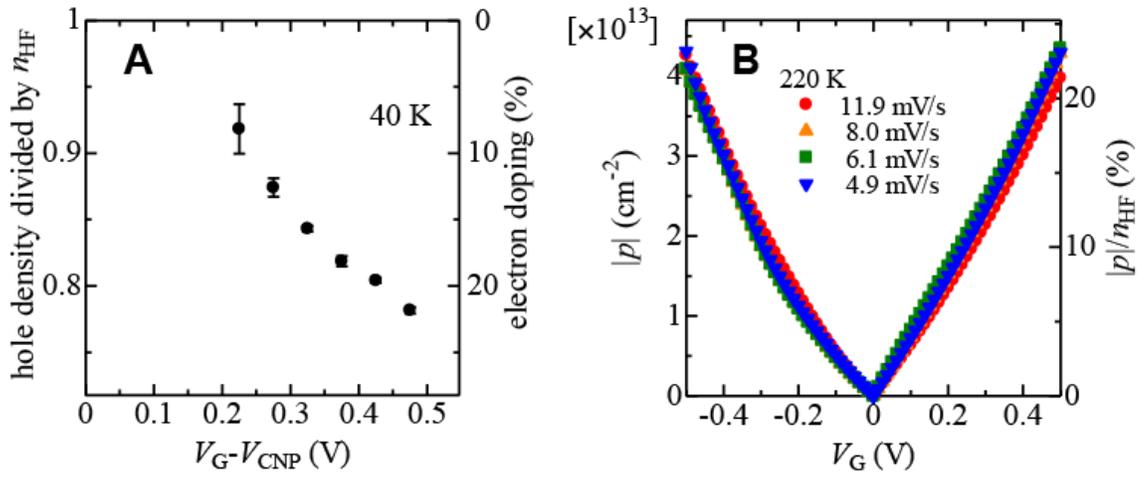

**Fig. S2.**
**Estimation of injected charge density.** (**A**) Gate voltage dependence of the hole density estimated from $1/eR_H$ in the Hall measurement. $V_{CNP}$ is the gate voltage of the charge neutrality point. (**B**) Gate voltage dependence of accumulated charge density $p$ in the displacement current measurement. The results for different sweep rates of $V_G$ are indicated by different symbols.

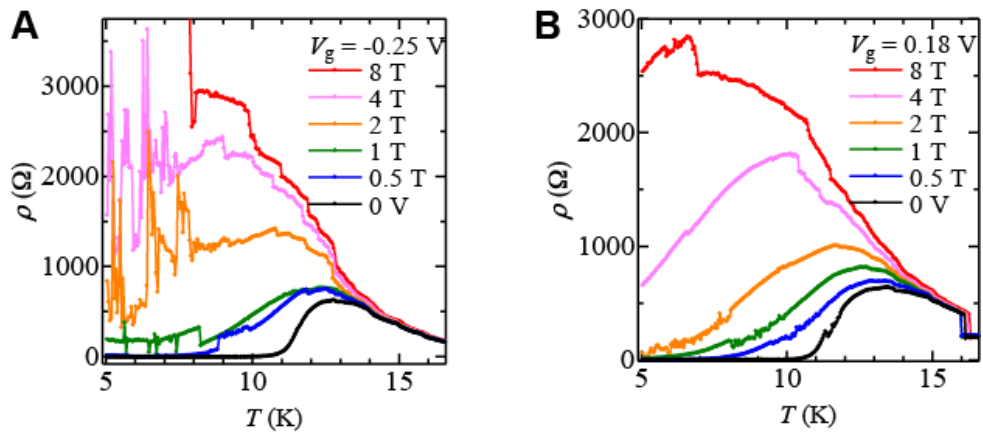

**Fig. S3.**
**Suppression of superconductivity by applying a magnetic field.** Temperature ($T$) dependence of the resistivity $\rho$ under different values of the applied magnetic field at $S = 0.39\%$ for (**A**) $V_G = -0.25$ V and (**B**) $V_G = 0.18$ V. The sample temperature was raised from 5 to 20 K.

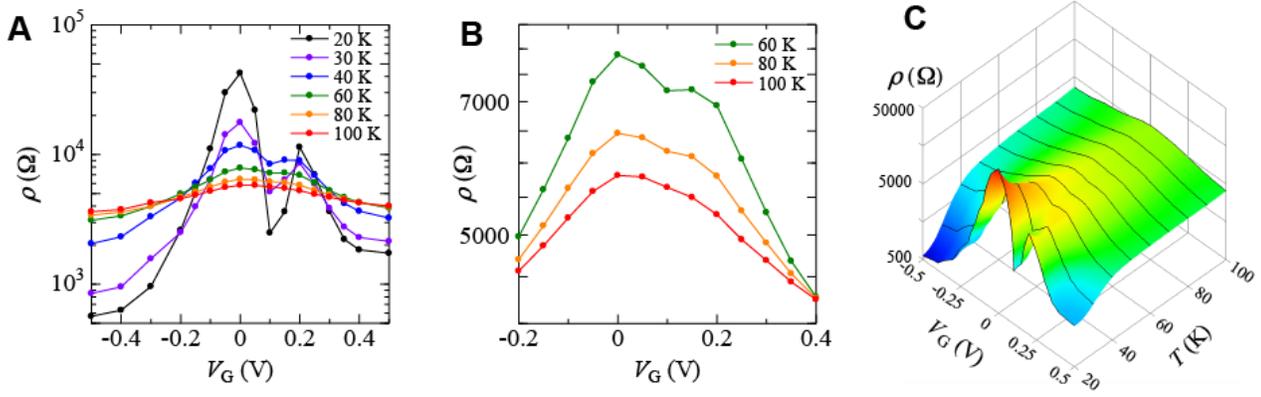

**Fig. S4.**
**Resistivity dip under small electron doping.** (**A**) Gate voltage dependence of the resistivity $\rho$ from 20 to 100 K. (**B**) Magnified plots from 60 to 100 K. (**C**) Three-dimensional plots of the data in A. Note that this sample is different from the one discussed in the main text.

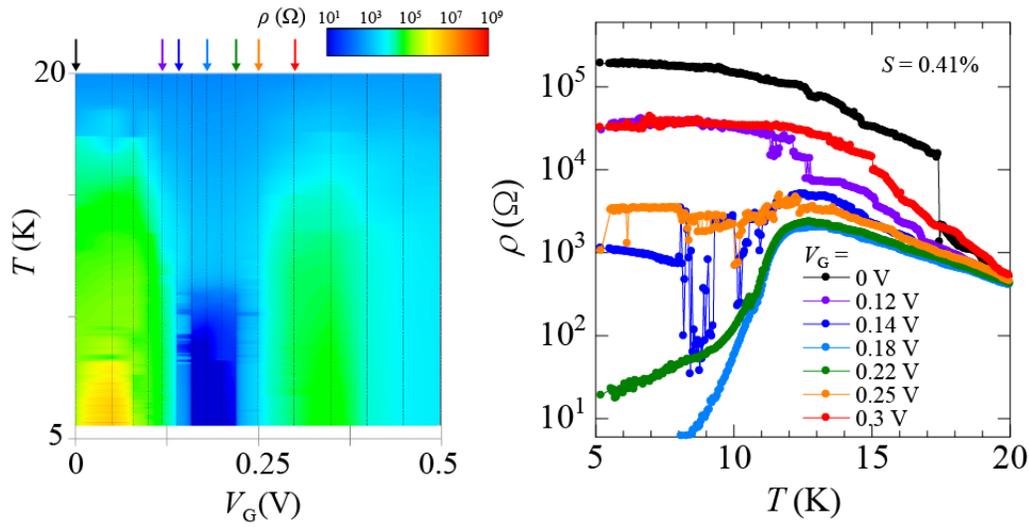

**Fig. S5.**
**Resistance vs temperature plots under low electron doping at tensile strain $S = 0.41\%$.** The left panel shows the electron-doped side of Fig. 2C and the arrows indicate the location of the corresponding data in the right panel. The resistivity fluctuates at the boundary between the insulating and superconducting states.

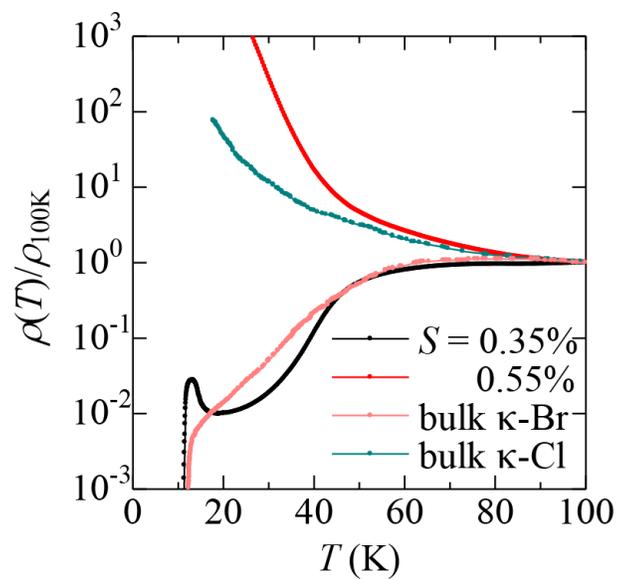

**Fig. S6.**
**Comparison of temperature dependences of the resistivity between our device and bulk crystals.** The data from bulk crystals are taken from Ref.(*3*).

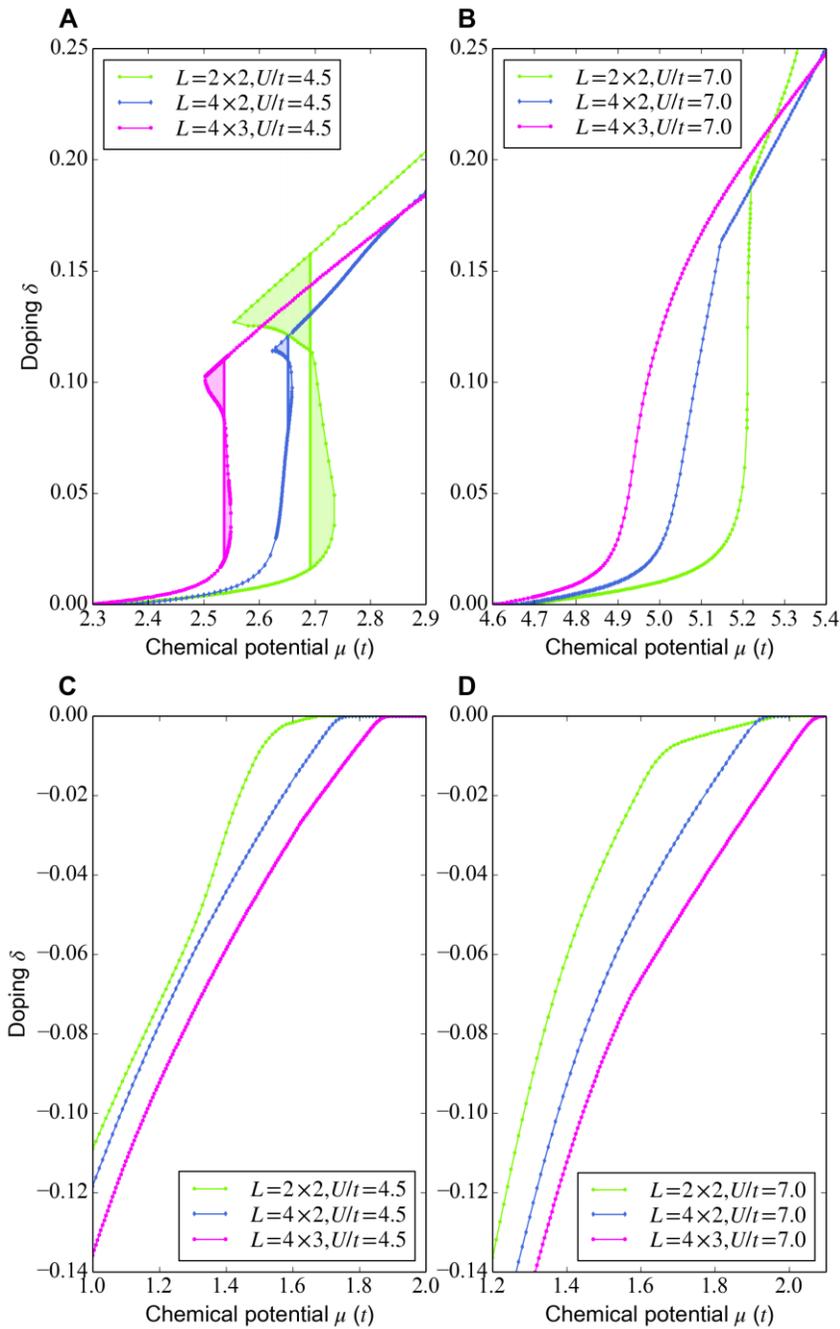

**Fig. S7.**
**Cluster size dependence of the phase separation tendency.** Doping concentration $\delta$ vs chemical potential $\mu$ for (**A**) $U/t = 4.5$ under electron doping, (**B**) $U/t = 7$ under electron doping, (**C**) $U/t = 4.5$ under hole doping, and (**D**) $U/t = 7$ under hole doping. The clusters of $L = 2 \times 2$, $L = 4 \times 2$, and $L = 4 \times 3$ sites are used. The nonmonotonic dependence of $\delta$ on $\mu$ in a, which should be replaced with the vertical straight line by the Maxwell construction, indicates the phase separation.

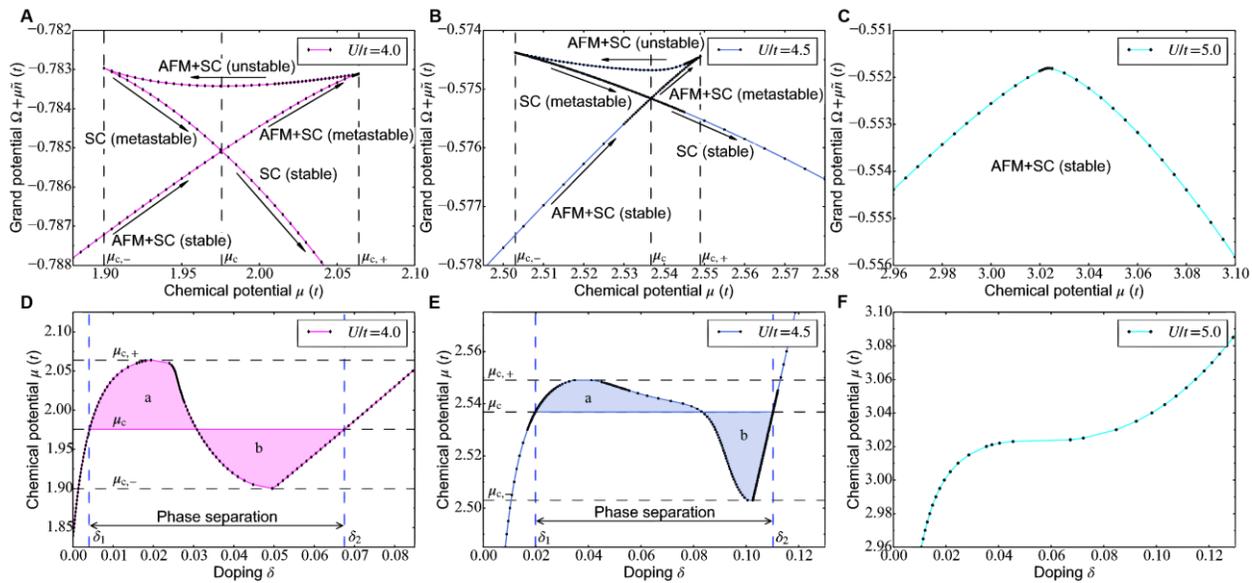

**Fig. S8.**
**Swallow-tail-shaped grand potential and Maxwell construction under electron doping with moderate correlation.** (**A**-**C**) Grand potential $\Omega$ vs chemical potential $\mu$ and (**D**-**F**) chemical potential $\mu$ vs doping concentration $\delta$ ($\mu$-$\delta$ curve) for (**A,D**) $U/t = 4$, (**B,E**) $U/t = 4.5$, and (**C,F**) $U/t = 5$. Notice that the linear term $\mu\tilde{n}$ with a constant (a) $\tilde{n} = 1.03075$, (b) $\tilde{n} = 1.0832$, and (C) $\tilde{n} = 1.06$ is added to the grand potential for clarity. The arrows in (**A**) and (**B**) indicate the direction along which $\delta$ increases. Therefore, the solutions associated with the leftwards arrows are unstable, for which $\delta$ increases with decreasing $\mu$ (see **D** and **E**). When the natural variable (or the experimentally controlling parameter) is $\mu$ as in the grand-canonical ensemble, the solutions associated with the rightwards arrows can be realized either as a stable or a metastable state, and the hysteresis might appear within the range of $\mu_{c,-} \leq \mu \leq \mu_{c,+}$ if $\mu$ is swept. However, the transition at the equilibrium should occur at $\mu = \mu_c$ indicated by dashed vertical lines. It is clearly observed in (**A**) and (**B**) that the grand potential exhibits the swallow-tail-shaped curve characteristic to the first-order transition, while in (**C**) the first-order transition no longer occurs. On the other hand, when the natural variable (or the experimentally controlling parameter) is $\delta$ as in the canonical ensemble, (**D**) and (**E**) show that the system may be separated into two phases with the doping concentrations $\delta_1$ (AFM+SC phase with smaller doping) and $\delta_2$ (pure SC phase with larger doping) indicated by the dashed vertical lines. The volume fraction of those two phases change with $\delta$. $\delta_1$ and $\delta_2$ in (**D**) and (**E**) are determined in order that the areas of the shaded regions (denoted by "a" and "b") are to be equal, according to the Maxwell construction. The monotonic behavior of $\mu$-$\delta$ curve in (**F**), although it is nearly flat around $\delta \approx 0.06$, indicates that the phase separation no longer occurs in this case. All the results are obtained for the single-band Hubbard model on the anisotropic triangular lattice ($t'/t = -0.44$) by using the VCA with the $4 \times 3$ cluster.